# Observation of the superconducting proximity effect in the surface state of $SmB_6$ thin films

**Seunghun Lee[1,2], Xiaohang Zhang[1,2], Yangang Liang[2], Sean Fackler[2], Jie Yong[1,3], Xiangfeng Wang[1,3], Johnpierre Paglione[1,3], Richard L. Greene[1,3], and Ichiro Takeuchi[1,2,*]**

[1]Center for Nanophysics and Advanced Materials, University of Maryland, College Park, Maryland 20742, USA

[2]Department of Materials Science and Engineering, University of Maryland, College Park, Maryland 20742, USA

[3]Department of Physics, University of Maryland, College Park, Maryland 20742, USA

*Corresponding author (email: takeuchi@umd.edu)

(Dated on April 5, 2016)

**ABSTRACT**

The proximity effect at the interface between a topological insulator (TI) and a superconductor is predicted to give rise to chiral topological superconductivity and Majorana fermion excitations. In most TIs studied to date, however, the conducting bulk states have overwhelmed the transport properties and precluded the investigation of the interplay of the topological surface state and Cooper pairs. Here, we demonstrate the superconducting proximity effect in the surface state of $SmB_6$ thin films which display bulk insulation at low temperatures. The Fermi velocity in the surface state deduced from the proximity effect is found to be as large as $10^5$ m/s, in good agreement with the value obtained from a separate transport measurement. We show that high transparency between the TI and a superconductor is crucial for the proximity effect. The finding here opens the door to investigation of exotic quantum phenomena using all-thin-film multilayers with high-transparency interfaces.

I. INTRODUCTION

Topological insulators (TI), which show exotic metallic surfaces with insulating bulk, have attracted tremendous interest in the condensed matter physics community. The topologically protected surface state of TI has been regarded as a promising platform for exploring exotic quantum phenomena in solid materials. In particular, it has been suggested that the superconducting proximity effect occurring at an interface between a TI and a superconductor may give rise to chiral topological superconductivity[1-3] and Majorana fermion excitations[4-9]. To this end, concerted experimental efforts have been made to study the superconducting proximity effect in superconductor/TI bilayer structures. However, in most Bi- and Te-based TI such as $Bi_2Se_3$, $Bi_2Te_3$, $Sb_2Te$, etc., the overwhelming conducting bulk electronic states hinder interfacial quantum phenomena and preclude the investigation of the interplay of the topological surface state and Cooper pairs[10-13]. In addition, low transparency at the superconductor/TI interface due to a non-pristine surface of the TI can significantly reduce the extent of the proximity effect[5,14-16]. Therefore, suppressing the bulk conductivity and securing high interfacial transparency in superconductor/TI bilayers have been an outstanding issue in this field.

Samarium hexaboride ($SmB_6$) has recently emerged as an ideal material to explore TI-based quantum phenomena because of its true insulating bulk[17]. To date, many theoretical and experimental studies have provided strong evidence for the existence of the conducting surface state and a robust insulating bulk of $SmB_6$ at low temperatures[18-37]. Specifically, electronic transport measurements on $SmB_6$ single crystals have revealed surface current dominated conduction[20,31] and thickness-independent Hall voltage at low temperatures due to the conducting surface[20]. A point contact spectroscopy (PCS) study has indicated the presence of an insulating bulk in $SmB_6$ at low temperatures[23,38]. The electronic structures of $SmB_6$ obtained by ARPES and a quantum oscillation measurement with torque magnetometry are consistent with theoretical predictions, *e.g.*, presence of an odd number of Dirac surface bands[19,34,39]. However, there still remain many questions regarding the nature of the surface state of $SmB_6$ including the origin of the metallic surface state and whether the surface state is truly topologically protected[17,22,27,40-42].

In this paper, we demonstrate direct observation of the proximity effect induced in the surface state of $SmB_6$ for the first time. We apply the Usadel formalism to the systematic change in the superconducting transition temperature of a series of *in-situ* deposited $SmB_6$/Nb bilayers with different Nb layer thicknesses and arrive at the normal coherence length of the surface state of $SmB_6$ to be ≈ 10 nm. From a separate transport study on $SmB_6$ thin films, we find the thickness of the surface state to be ≈ 7 nm. The Fermi velocity ($v_F$) obtained from the transport measurements is in good agreement with $v_F$ calculated using the normal coherence length, and it is of the order of $10^5$ m/s. These numbers paint a picture of a material whose bulk insulation provides a unique opportunity to probing and exploiting its surface state in thin-film multilayer devices.

## II. METHODS

$SmB_6$ thin films were prepared by co-sputtering of $SmB_6$ and B targets; additional B sputtering was performed for compensating B deficiency and the base pressure was ~ $2×10^{-8}$ Torr. The thin films were grown on Si (100) substrate at 800°C in an atmosphere of Ar (99.999%) and the working pressure was kept at 10 mTorr. After deposition, further annealing process was performed at 800˚C for 3 hours in high vacuum. X-ray diffraction and transmission electron microscopy results can be found in the reference[37]. For fabricating Nb/SmB$_6$ bilayers, Nb layer were prepared on SmB$_6$ by sputtering method both *in-situ* and *ex-situ* at room temperature. The base pressure was ~ $2\times10^{-8}$ Torr and the working pressure was kept at 7 mTorr. For fabricating Nb/Au bilayer, Au layer was prepared on Si substrate by thermal evaporator and the base pressure was ~ $10^{-6}$ Torr.

The electrical properties were measured by Physical properties measurement system (PPMS). Temperature-dependent resistance for $T_c$ evaluation was measured by simple 4-point probe method with 0.01 K step. Electrical contacts were made with an Al wire-bonder, and the typical contact resistance was ~ 1 Ω. Transport characteristics of SmB$_6$ thin films were measured with Hall bar geometry; six-contact 1-2-2-1 Hall bars with 200 μm channel size were prepared by ion milling process.

## III. RESULTS AND DISCUSSION

### A. Transport characteristics

We performed transport measurements on SmB$_6$ thin films with varying film thickness. In order to perform the transport measurements, each film was patterned into a Hall-bar with a channel width of 200 μm. Resistance vs. temperature (R-T) was measured for all SmB$_6$ films, and their sheet conductance ($G_\square$) values at 300 K and 2 K are plotted in Figure 1. The $G$ was calculated using $G_\square = G \cdot a/w$, where $a$ and $w$ are the length and the width of the Hall bar channels, respectively. All samples show a nearly constant resistance value at low temperatures. The inset shows a representative R-T curve for a 25 nm thin film. The result is consistent with previous reports[24,29,31,35] where a saturation of resistance (*i.e.*, plateau) has been attributed to the transition from bulk state-dominated conduction to surface state-dominated conduction at low temperatures.

To see if there is any thickness dependent effect in our samples, we plot $G$ at 2 K and 300 K as a function of film thickness (Figure 1). $G$ at 300 K clearly displays a linear thickness dependence, indicating bulk transport. However, if the resistance plateau at low temperatures arises from the surface conduction, $G$ at 2 K should be independent of the thickness and become a constant. This is indeed the case, as seen in Fig. 1, strongly suggestive of the presence of the surface conduction channel. We also measured the Hall resistance ($R_{xy}$) as a function of magnetic field, and calculated the sheet carrier concentration and the mobility of the $SmB_6$ thin films at 2 K. Both were found to be constant and independent of the film thickness (Supplementary Info.).

To determine the thickness of the surface conduction channel, we adopt a simple parallel conductance model where the electronic conduction is through two channels: a surface channel including contributions from both the top and the bottom surfaces and a bulk channel[24,31,43]. This model is valid for the temperature region where the Kondo gap of $SmB_6$ opens, which is a sufficient condition for the emergence of the topological surface state. Therefore, we apply this model to the transport results in a temperature range below 120 K, at which the Kondo gap starts to open in $SmB_6$[23]. The conductance of the surface channel is assumed to be temperature independent, and that of the bulk channel is temperature dependent and it can be described by a bulk resistivity and exponential function (*i.e.,* Arrhenius equation). The total conductance of the $SmB_6$ thin film can then be described as:

$$G = G_{surface} + G_{bulk} \quad (1)$$

$$G_{surface} = \frac{w}{a} G_{,2K}, \quad G_{bulk} = \frac{w \cdot t_{bulk}}{a} \cdot \sigma_{bulk,300K} \exp\left(-\frac{E_a}{k_B T} + \frac{E_a}{300 k_B}\right) \qquad (2)$$

The use of the bulk conductivity in the $G_{bulk}$ term allows us to add a thickness of bulk channel ($t_{bulk}$) as fitting parameters, and subsequently to estimate a thickness of surface channel ($t_{surface}$) because the total film thickness is $2t_{surface} + t_{bulk}$. The bulk conductivity, $\sigma_{bulk,300K}$, is $7.66 \times 10^5$ S m$^{-1}$ which is the slope of $G$ at 300 K obtained from the data shown in Fig. 1. $G_{,2K}$ is the $G$ at 2 K for each sample, and $a$ and $w$ are the length and the width of the Hall bar channel, respectively. Fitting parameters are $t_{bulk}$ and $E_a$, where $E_a$ is an activation energy. We fit the data for all samples using the equation (1) and (2). A representative fitting result is shown in Fig. 2a. The values of fitting parameters $E_a$ and $t_{surface}$ for the different thickness samples are shown in Figure 2b, and $t_{surface}$ is the thickness of the surface conduction channel, *i.e.*, the thickness of the surface state of SmB$_6$. Both $E_a$ and $t_{surface}$ are found to be constant regardless of the film thickness. The average value of $E_a$ is $2.73 \pm 0.04$ meV, which is in good agreement with the values obtained from SmB$_6$ single crystals[24,28,44].

From the above analysis, we obtain $t_{surface} \approx 7$ nm. To obtain $v_F$, we used $v_F = E_g t_{surface}/\hbar$ derived from the solution of the effective low-energy Hamlitonian for topological insulators with a surface state where $E_g$ is the bulk energy gap[29,45]. We use $E_g$ of 17 meV measured on our SmB$_6$ thin films by terahertz spectroscopy[46], which is also in agreement with the values obtained from single crystal SmB$_6$ using point contact spectroscopy[23], scanning tunneling spectroscopy[25], etc. With $E_g = 17$ meV and $t_{surface} = 7$ nm, we arrive at $v_F \sim 1.8 \times 10^5$ m/s, a value close to that measured in quantum oscillation measurements on single crystal SmB$_6$[19]. This value is also in agreement with an independent estimate of $v_F$ obtained from the proximity effect study as discussed below.

**B. Proximity effect**

The superconducting proximity effect describes a phenomenon at a superconductor-normal metal interface where Cooper pairs diffuse into the normal metal resulting in the suppression of the critical temperature ($T_c$) of the superconductor while inducing surface or local superconductivity in the normal metal. We fabricated Nb/SmB$_6$ bilayers and observed a change in $T_c$ depending on the thickness of Nb layer due to the proximity effect at the interface. To characterize the proximity effect of the bilayers, we treat the Nb/SmB$_6$ bilayer as a superconductor/metal bilayer system, where the metallic layer in SmB$_6$ is the surface conducting channel (Figure 3), and we calculate the normal coherence length ($\xi$) and $v_F$ of SmB$_6$ thin films using the Usadel equation[16,47,48]. Assuming that $\xi$ of SmB$_6$ ($\xi_{SmB_6}$) is longer than the thickness of the surface conducting channel of SmB$_6$ ($\xi_{SmB_6} > t_{surface}$), the fitting equation for evaluating $\xi_{SmB_6}$ can be obtained by linearizing the Usadel equation[16]:

$$\frac{T_{cb}}{T_{cs}} = 1 - \frac{\pi^2}{4} \frac{\xi_{Nb}}{d_{Nb}} \gamma \cdot p(\gamma) \qquad (3)$$

$$\gamma = \frac{\rho_{Nb}\xi_{Nb}t_{surface}}{\rho_{SmB_6}(\xi_{SmB_6})^2} \qquad (4)$$

$$p(\gamma) = 1.17 + \frac{2}{\pi^2}\ln(1 + 0.98\gamma^{-2}) \qquad (5)$$

where $T_{cb}$ and $T_{cs}$ represent the $T_c$ of the Nb/SmB$_6$ bilayer and the $T_c$ of a single Nb layer, respectively. $T_{cb}$ is evaluated by the Ginzburg-Landau equation, and $\gamma$ and $p(\gamma)$ are approximated by the expressions above[16]. $\gamma$ represents the strength of the proximity effect between layers[47]. $\rho_{Nb}$ and $\rho_{SmB_6}$ are the residual resistivities of the Nb layer and the SmB$_6$ layer, respectively. $\xi_{Nb}$ and $\xi_{SmB_6}$ are the coherence length of the Nb layer and the SmB$_6$ layer, respectively. $d_{Nb}$ is the thickness of the Nb layer. To check the validity of the model, a series of Nb/Au bilayers were also fabricated, and the extracted value of $\xi_{Au}$ was in good agreement with their known value (~ 1μm).

For Nb/SmB$_6$ bilayers, $d_{Nb}$ was varied from 20 to 100 nm, and the thickness of SmB$_6$ layer was fixed to be 50 nm. Figure 4a and 4b show the resistance vs. temperature curves of single Nb layers and Nb/SmB$_6$ bilayers near $T_c$ for different thicknesses of the Nb layer ($d_{Nb}$) respectively. Each resistance curve was normalized by a value obtained at a temperature slightly above the transition temperature. The $T_c$ of single Nb layer gradually decreases with decreased $d_{Nb}$ due to localization[49] and each Nb/SmB$_6$ bilayer has a lower $T_c$ than the corresponding Nb single layer due to the proximity effect established between Nb and SmB$_6$ layers. Based on x-ray diffraction and cross-sectional SEM of the bilayers, we do not believe there is any significant diffusion at the interface.

Figure 4c shows the $T_{cb}/T_{cs}$ of the Nb/SmB$_6$ system as a function of $d_{Nb}$. As a comparison, one bilayer was made with an *ex-situ* interface: following the deposition of SmB$_6$, it was exposed to air before the Nb layer was deposited. As represented in Fig. 4c, this sample shows a much higher $T_{cb}/T_{cs}$ (*i.e.*, higher $T_c$) than the corresponding *in-situ* bilayer. The proximity effect is very sensitive to the nature of the top most layer of SmB$_6$. Specifically, degradation of the top most surface layer (*e.g.*, due to oxidation) may reduce the boundary transparency, resulting in reduction of the proximity effect. This result demonstrates that *in-situ* formed multilayers with clean interfaces are critical for establishing such a proximity effect at the interface of SmB$_6$.

To evaluate $\xi_{SmB_6}$, we performed a fitting process based on the expression discussed above. Using measured values of $\rho_{Nb} = 8.5 \times 10^{-6}$ Ω·cm, the mean free path (*l*) of Nb was estimated from $\rho_{Nb}l = 3.75 \times 10^{-6}$ μΩ·cm² [50] (*i.e.*, $l$ = 4.4 nm) and we obtain $\xi_{Nb} = 0.852(\xi_{Nb(bulk)}l)^{1/2}$ = 11 nm where $\xi_{Nb(bulk)}$ is the known value of coherence length of Nb bulk ($\xi_{Nb(bulk)}$ = 38 nm)[51]. The thickness of the surface conducting channel of $SmB_6$ has been evaluated from our transport result ($t_{surface}$ = 7 nm), and the resistivity of the surface state was calculated to be $8.96\times10^{-5}$ Ω·cm (= $\rho_{SmB_6}$) at 2 K based on the above measured sheet conductance ($\rho=2t_{surface}/G$ ). The only unknown parameter is $\xi_{SmB_6}$. The fitting result is shown as a red solid line in Fig. 4c, and we arrive at $\xi_{SmB_6}$ of 9.6 nm. To calculate $v_F$, $\xi=(v_F\hbar l/6\pi k_B T)^{1/2}$ is used[52], where $l$ of the $SmB_6$ thin film was assumed to be the grain size of our $SmB_6$ film, ≈ 4 nm[37], and we obtain $v_F \approx 10^5$ m/s at 2 K for $SmB_6$. This $v_F$ is comparable to the value obtained from the transport study above ($v_F = 1.8\times10^5$ m/s), which implies that the observed superconducting proximity effect is attributed to the surface state of $SmB_6$ thin film. As a comparison, we have also carried out a fit using entire bulk of the $SmB_6$ thin film (even though, we have shown above that the bulk of the film is insulating): in this case, the fit does not provide a $v_F$ value consistent with the $v_F$ value obtained from the transport study. Our bilayer fabrication method in ultra high vacuum process excludes any degradation or contamination as the origin of the metallic surface state.

The value of $v_F$ and the effective mass of quasiparticles in the surface state of $SmB_6$ have been the subjects of much debate. Early theoretical calculations had suggested the presence of relatively heavy Dirac quasipaticles with small $v_F$ (~ $10^3$ m/s)[26,53] in the topologically protected surface state, while recent APRES and quantum oscillation studies have pointed to light qusiparticles with large $v_F$ of > $10^4$ m/s[19],[32,33]. To resolve the discrepancy, Alexandrov *et al.* have recently proposed that Kondo breakdown at the surface of $SmB_6$ could release unquenched moments at the surface, causing the Dirac point to shift down into the valence band and giving rise to large $v_F$ values[54]. We have examined the Sm valence state in the surface of our $SmB_6$ thin films using x-ray photoemission spectroscopy at room temperature and found that it is similar to that of $SmB_6$ single crystals, which is ~ 2.7[55] (Supplementary Info.). Therefore, we exclude chemical extrinsic effects as the origin of the large $v_F$ value observed here. We believe our high $v_F$ value is also due to the Kondo breakdown effect[54]. We note that $t_{surface}$ of ≈ 7 nm found in this study is consistent with the thickness predicted with the Kondo breakdown effect[54]. It has been reported that $SmB_6$ has three Dirac cones, each with its own slightly different value of $v_F$[32,39] The measured $v_F$ in this study is therefore expected to be the average $v_F$ of the three Dirac cones.

With the $\xi_{SmB_6}$ obtained from the fitting above, we are able to evaluate $\gamma$ which is a measure of the strength of the superconducting proximity effect. The values of $\gamma$ for *in-situ* and *ex-situ* bilayer are $7.9 \times 10^{-2}$ and $1.0 \times 10^{-2}$, respectively. This implies that the strength of the proximity effect of the *in-situ* sample is roughly 8 times larger than that of the *ex-situ* sample. The strength of the superconducting proximity effect is naturally one of the most important factors in superconductor – normal– superconductor Josephson junctions, and it directly affects the $I_C R_N$ product. It is interesting to note that significantly reduced $I_C R_N$ products, presumably due to lack of the pristine surface of the TI during the fabrication process, have been observed in various studies on chalcogenide-TI-based Josephson junctions[5,6]. Specifically, the observed $I_C R_N$ products in such studies are typically around 20 μV, which is far below the theoretical value, *i.e.*, $I_C R_N \sim \pi \Delta(0)/2e \approx 4.7$ mV, for junctions in the clean limit [56][57]. Higher $I_C R_N$ product junctions (and in particular high $I_C$ junctions) are always desirable for investigating novel quantum phenomena involving coherence of superconductivity, where signature of such phenomena might appear in the form of small modulation in the critical current. The present study shows that the strength of proximity effect in the bilayer with an *in-situ* interface is nearly an order of magnitude higher than that with an *ex-situ* interface. We expect that with identical geometric dimensions, junctions fabricated by *in-situ* interface may display a substantially larger $I_C R_N$ than junctions fabricated with *ex-situ* interface. Thus, we have demonstrated an important pre-requisite for attaining a high $I_C R_N$ junction. A non-ideal S/N interface such as one with a thin tunnel barrier (formed due to an *ex-situ* process) can result in scattering at the interface leading to decoherence[7,58]. This may be the reason why the proposed signatures of Majorana fermions have not been observed in Josephson junctions to date.

## IV. CONCLUSIONS

In summary, thickness-independent behavior in the transport properties was observed in ultrathin $SmB_6$ films, and the thickness of the surface state was deduced to be ≈ 7 nm. We provide first direct evidence of the superconducting proximity effect in the surface state of $SmB_6$ through pristine interfaces with *in-situ* deposited Nb layers. The Fermi velocity values of the surface state obtained from the transport measurements and the proximity effect are in good agreement with each other. The present work lays the groundwork for fabricating $SmB_6$ thin film based multilayers and devices for investigating quantum phenomena including Majorana fermion excitations.

## ACKNOWEDGEMENT

The authors would like to acknowledge professor Victor Galitski for valuable discussions. This work was supported by AFOSR (FA95501410332) and NSF grants (DMR-1410665 and DMR-1410665). It was also supported by the Maryland NanoCenter.

**Figures and captions**

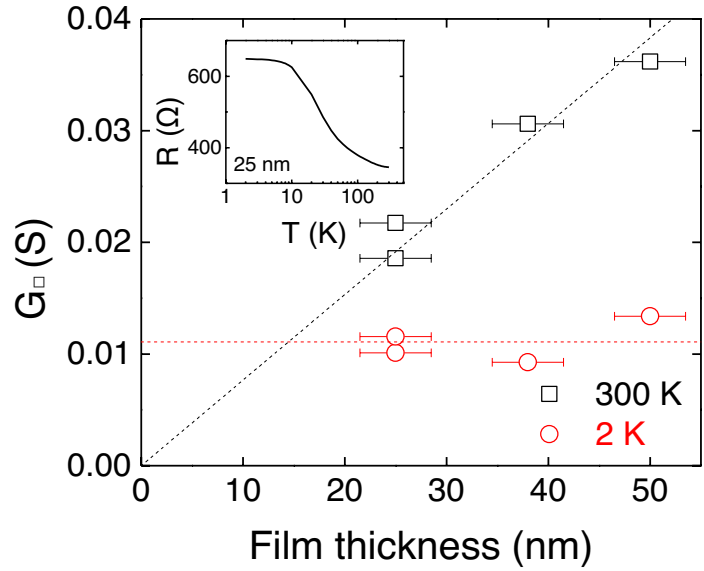

**Figure 1. Sheet conductance ($G_\square$) of SmB$_6$ thin films at 2 K and 300 K as a function of thin thickness** Dashed lines are linear fits to the experimental results. The slope of 300 K data represents 3D bulk conductivity at 300 K in SmB$_6$ thin films, and the fit to the data (black dot line) extrapolates to zero at zero thickness, consistent with the fact that the conduction in SmB$_6$ thin films is dominated by a 3D bulk transport at 300 K. At 2 K, $G_\square$ is independent to the thickness, and the guide line (red dot line) is the fit to the data, which indicates the average value of $G_\square$ at 2 K. The inset shows resistance vs. temperature curve for a 25 nm SmB$_6$ thin film.

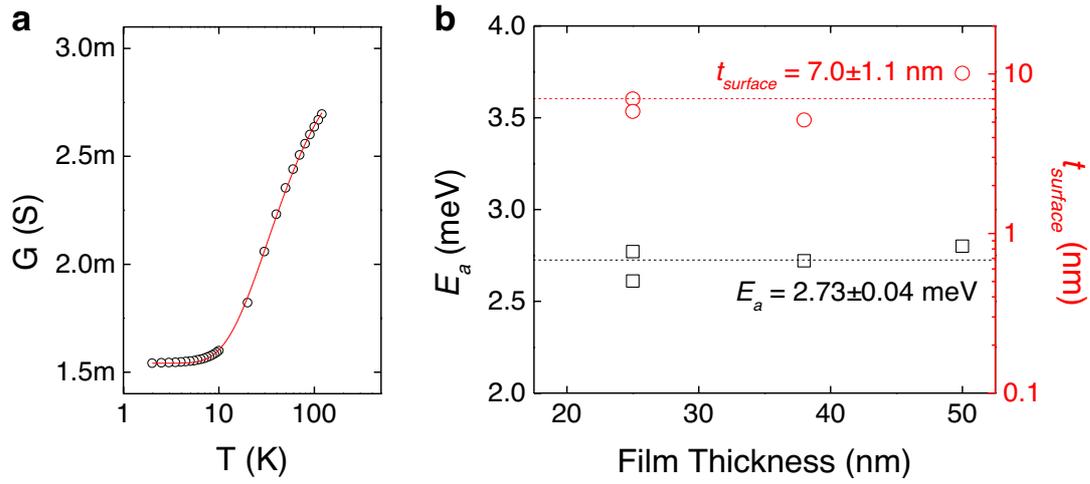

**Figure 2. Estimating the thickness of the surface conduction channel in SmB$_6$ thin films** (a) Conductance vs. temperature of a 25 nm SmB$_6$ thin film. The red line is a best fit to the experimental data using the parallel conductance model. (**b**) Activation energy ($E_a$) and the thickness of the surface conduction channel ($t_{surface}$) extracted from the fit as a function of the film thickness for SmB$_6$.

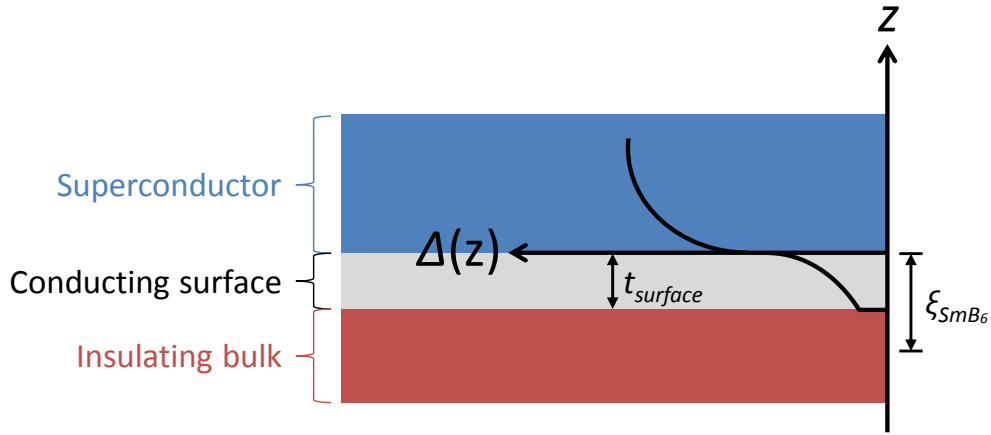

**Figure 3. Model for the proximity effect for superconducting/TI system** Given the surface conducting channel of the $SmB_6$ as discussed above, we adopt the model of the proximity effect for the superconducting layer and the surface conduction channel, the surface state of $SmB_6$. The plot in the schematic represents the evolution of the superconducting pair potential ($\Delta(z)$). In this model, we consider the case where $\xi$ of $SmB_6$ ($\xi_{SmB_6}$) is larger than the thickness of the surface state of $SmB_6$ ($t_{surface}$).

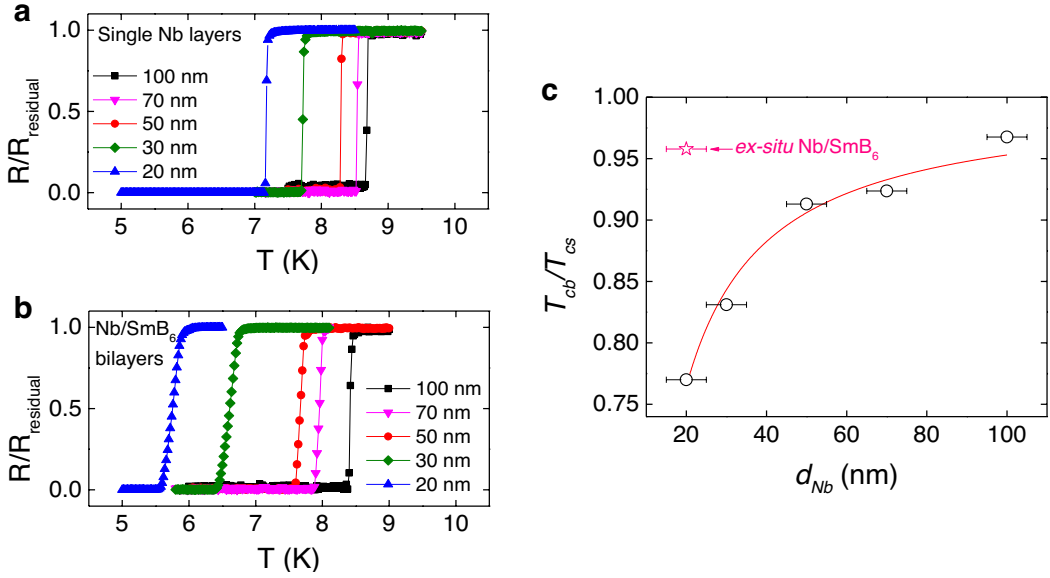

**Figure 4. Proximity effect in Nb/SmB$_6$ bilayer.** Resistance vs. temperature curves of (**a**) single Nb layers and (**b**) Nb/SmB$_6$ bilayers for different thickness of Nb layers ($d_{Nb}$) where the thickness of the SmB$_6$ layer is fixed to be 50 nm. Each Nb layer was deposited on SmB$_6$ *in-situ*. The resistance values are normalized by a value obtained at a temperature slightly above the transition temperature (R$_{residual}$). (**c**) $T_{cb}/T_{cs}$ (*i.e.*, ratio of $T_c$ of Nb/SmB$_6$ bilayer to $T_c$ of single Nb layer) as a function of $d_{Nb}$. The solid line is a fit using the Usadel equation. The star symbol in Fig. 4c indicates $T_{cb}/T_{cs}$ of the Nb/SmB$_6$ bilayer where 20 nm Nb was deposited on SmB$_6$ after its surface was first exposed to air forming an *ex-situ* interface with low transparency.

# Supplementary Information

**- Thickness independent carrier concentration and mobility**

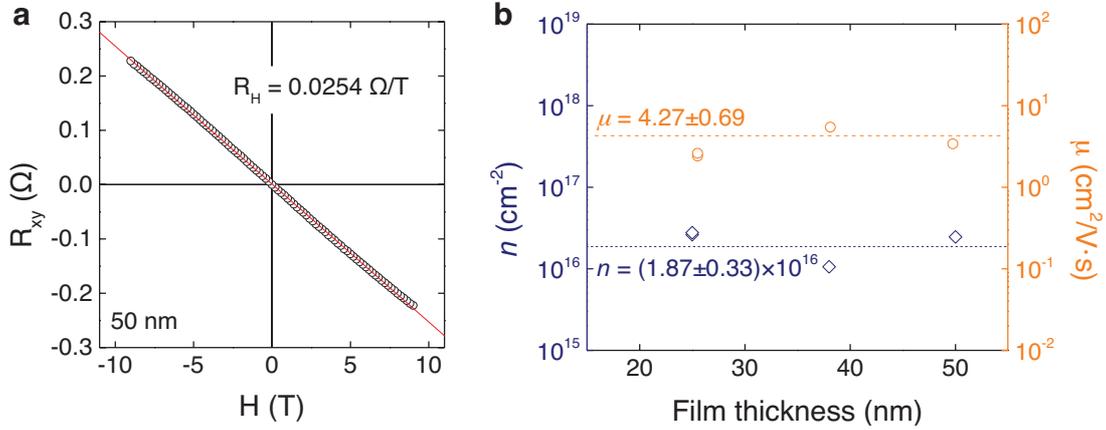

**Figure S1. Transport characteristics of SmB$_6$ thin films.** (**a**) Hall resistivity ($R_{xy}$) vs. field measured at 2 K for a 50 nm SmB$_6$ film. (**b**) Sheet carrier concentration ($n$) and mobility ($\mu$) for different thickness SmB$_6$ films.

To further investigate the surface state, we measured the Hall resistance ($R_{xy}$) as a function of magnetic field, and calculated the carrier concentration and the mobility of the SmB$_6$ thin films. As shown in Figure S1a, the Hall resistivity $R_{xy}$ of the 50 nm film appears to follow a straight line at 2 K in the entire magnetic field range we studied, suggesting that only one type of carriers contributes the electrical conductivity at 2 K and that bulk contribution is well suppressed: if the bulk contribution persists at such low temperatures, we would expect there to be multiple types of carriers with different concentrations and mobility values, leading to a nonlinear field dependence of $R_{xy}$ as it has been previously observed in other TIs such as Bi$_2$Se$_3$[1,2]. In contrast, the linear field dependence we observed indicates a presence of a single carrier in SmB$_6$ thin films at 2 K and the insulating bulk state attributed to the Kondo insulating nature.

The slope of $R_{xy}(B)$ is determined by $1/ne$ where $n$ is the sheet carrier concentration. A mobility, $\mu$ of carriers can be calculated from $\mu = G_\Box/en$. We obtain both $n$ and $\mu$ for each thin film and plot them as a function of film thickness in Figure S1b. Both $n$ and $\mu$ appear to be constant regardless of the variation in the film thickness. The average values of $n$ and $\mu$ are $1.87 \times 10^{16}$ cm$^{-2}$ and 4.27 cm$^2$/Vs, respectively. As expected from the observed behavior of $G_\Box$, because the surface conduction is dominant at low temperatures, the transport parameters ($n$ and $\mu$) are also independent of the thickness as seen in Fig. S1b.

- **Valence state of the Sm ion near surface**

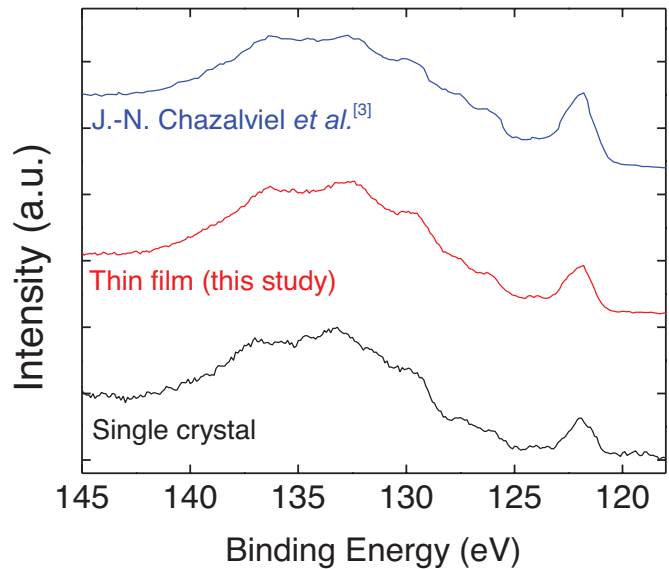

**Figure S2. Investigating the valence state of Sm ion in a SmB$_6$ thin film** Comparison of x-ray photoelectron spectroscopy (XPS) Sm 4$d$ spectra of a SmB$_6$ thin film, a single crystal SmB$_6$ (measured) and bulk[3] (reference)

To investigate the valence state of Sm ions, we used x-ray photoelectron spectroscopy (XPS) to measure the Sm 4$d$ spectrum of a SmB$_6$ thin film. We also measured the Sm 4$d$ spectrum of a SmB$_6$ single crystal as a reference. Details on the single crystal growth and properties can be found elsewhere[4,5]. The Sm 4$d$ spectra of the SmB$_6$ thin film and the single crystal are shown in Fig. S2: the Sm 4$d$ spectrum consists of Sm$^{2+}$ and Sm$^{3+}$ related multiplet structures at lower and higher energies than the binding energy of ~ 130 eV, respectively, which partially overlap. As shown in Fig. S2, the Sm 4$d$ spectra of the single crystal and the thin film are almost identical, implying that the Sm valence state in the thin film and the single crystal is the same. For comparison, we also included the reported Sm 4$d$ spectrum[3] of a bulk SmB$_6$ sample showing the mixed valence state in the ratio Sm$^{2+}$:Sm$^{3+}$ ≈ 1:3 in Fig. S2. This looks very similar to both the spectra of the single crystal and the thin film. This indicates that the Sm valence at surfaces of both the thin film and the single crystal is similar to the bulk valence state of $SmB_6$ (≈ 2.7)[6].